\documentclass[prb,twocolumn,showpacs,floatfix,superscriptaddress]{revtex4}
\usepackage{eepic}
\usepackage{epsf}
\usepackage{amsmath}
\newcounter{saveeqn}%

\begin{document}
\title{Prediction of infrared light emission from $\pi$-conjugated polymers: a 
diagrammatic
exciton basis valence bond theory}
\author{S. Dallakyan}
\affiliation{ Department of Physics, University of Arizona
Tucson, AZ 85721}
\author{M. Chandross}
\affiliation{Sandia National Laboratories, Albuquerque, NM 87185-1411}
\author{S. Mazumdar}
\affiliation{Department of Physics and The Optical Sciences Center, 
University of Arizona, Tucson, AZ 85721}.
\date{\today}
\begin{abstract}
There is currently a great need for solid state lasers that emit in 
the infrared, as this is the operating wavelength regime for applications in
telecommunications. Existing $\pi$--conjugated polymers all emit in the 
visible or ultraviolet, and whether or not $\pi$--conjugated polymers that emit in the infrared
can be designed is an interesting challenge. 
On the one hand,
the excited state ordering in trans-polyacetylene, the $\pi$--conjugated 
polymer with relatively small optical gap,
is not conducive to light emission because of electron-electron interaction
effects. On the other hand, excited state ordering opposite to that in 
trans-polyacetylene
is usually obtained by chemical modification that increases the effective 
bond-alternation, which in turn increases the optical gap. 
We develop a theory of electron correlation effects in a model
$\pi$-conjugated polymer that is obtained by replacing the hydrogen atoms
of trans-polyacetylene with transverse conjugated groups, and show that the
effective on-site correlation in this system is smaller than the bare
correlation in the
unsubstituted system. An optical gap in the infrared as well as excited state
ordering conducive to light emission is thereby predicted upon similar 
structural modifications.
\end{abstract}
\maketitle
\section{Introduction}
\label{intro}
$\pi$-conjugated polymers have attracted wide attention in the past decade
as light emitting organic semiconductors.
Light emitting diodes with $\pi$-conjugated polymers as emissive 
materials have
reached very high efficiencies and lifetimes \cite{Friend}, 
and there has also been 
considerable activity in the area of lasing using organic
materials since the original demonstration of spectral narrowing in these
systems under intense optical pumping \cite{Tessler,Hide,Frolov}.
One serious limitation in this area has been, however,
that all light emitting $\pi$-conjugated polymers
to date emit in the visible or UV. Telecommunications use infrared
radiation, so lasing at these wavelengths is desirable \cite{PT}.
Whether or not 
$\pi$-conjugated polymers can be synthesized that have light emission in the
IR, and what should be the structural features of such a polymer are
interesting theoretical questions. Theoretical work on these topics was
recently initiated by one of us and his collaborators \cite{Shukla,Ghosh},
where however, the emphasis was on explaining the unexpectedly high
experimentally observed photoluminescence (PL) efficiency in 
polydiphenylacetylenes (PDPA)
\cite{Tada,Gontia,Hidayat}, which consist of a backbone carbon chain that is 
the same as in nonemissive polyenes and side groups that are phenyl groups
instead of atomic hydrogen.
It was claimed that in an ideal long
chain of PDPA the optical gap is smaller than that in trans-polyacetylene
(t-PA). Real PDPAs emit in the visible because of their
finite conjugation lengths \cite{Tada}, but it was claimed in the earlier
theoretical work \cite{Shukla,Ghosh} that relatively small optical
gaps and strong PL would be
expected in any 
$\pi$-conjugated polymer where in addition to longitudinal conjugation there
occurs conjugation over a few molecular bonds in the transverse direction.

PDPA's, however, because of their complicated electronic structures cannot
be considered as the prototype of these novel class of materials. Furthermore,
although the previous calculations \cite{Shukla,Ghosh} were carried out 
to a high level of accuracy using the multiple reference doubles configuration
interaction (MRDCI) technique, by themselves they do not give a 
clear mechanistic explanation of {\it why}
the ordering of the excited energy states
in PDPA's are opposite to that in the linear polyenes such that light emission
becomes feasible
In the present paper we therefore
choose the simplest model system that shares the structural feature of
transverse conjugation with PDPA's,
and analyse its lowest excitations. We demonstrate
that finite transverse conjugation over a few bonds
is the key to obtaining light emission in
the IR (provided, of course, long chain systems can be synthesized). Our
theoretical work shows that there exists a different mechanism for the
energy crossover between the lowest one- and two-photon states in
$\pi$-conjugated polymers than the more common one that depends on
effective bond alternation \cite{Soos1,Soos2}. The simplicity of our model
system allows us to perform numerical work 
using a diagrammatic exciton basis valence bond method
\cite{Chandross} within which the basis functions retain their local
character, and diagrammatic interpretations of wavefunctions can be given. 
A complete insight to the theory of excited state ordering in these complicated
systems is thus obtained, which at the same time fully supports the conclusions
of our earlier work \cite{Shukla,Ghosh} that used a more traditional approach.
To the best of our knowledge, although the exciton basis has 
often been used in the past for calculations on 
coupled two-level systems (often though
with severe approximations) 
\cite{Simpson,Ohmine,Ishida,Gallagher,Mukho,Barford,Lavrentiev,Pleutin},
there exist few
such calculations for coupled multilevel systems 
\cite{Rice,Chakrabarti,Spano}, again sometimes
within simpler models. Additional interest of our work then comes from
its applicability to realistic multilevel systems.

Emission in the IR would require $\pi$-conjugated polymers whose optical gaps
are smaller than or comparable to that of t-PA.
Linear polyenes and t-PA are, however, 
weakly emissive, because the lowest two-photon
state, the 2A$_g$, occurs below the optical 1B$_u$ state in these 
\cite{Hudson}. The same is also true for long-chain polydiacetylenes (PDAs).
\cite{Lawrence}. This is a conseqence of moderate 
electron-electron (e-e) interactions between the $\pi$-electrons, and is
by now well understood \cite{Ohmine,Hudson,Ramasesha,Shuai,Lavrentiev1}.
The optically pumped 1B$_u$ in these systems decays to 
the 2A$_g$ in ultrafast times, and
radiative transition from the 2A$_g$ to the ground state 1A$_g$ is forbidden.
Strong PL in systems like poly-paraphenylenevinylene (PPV) and 
poly-paraphenylene (PPP) implies
excited state ordering E(2A$_g$) $>$ E(1B$_u$) [where E(...) is the energy of
the state], which is a consequence of enhanced bond alternation
within the {\it effective} linear chain model for
these systems \cite{Soos1,Soos2}. Within the effective model, the lowest 
excitations of systems containing phenyl rings as part of the conjugated main
chain can be mapped onto those of a
linear chain with bond alternation that is much larger than that in t-PA.
Explicit calculations then indicate that even with the same e-e interactions,
chains with larger bond alternation can have E(2A$_g$) $>$ E(1B$_u$).
Since enhanced bond alternation necessarily
{\it increases} E(1B$_u$), it appears that strong PL should then be limited
to systems with optical gaps larger than that of t-PA, and emission in the
IR from $\pi$-conjugated polymers would be impossible.

Our goal is to demonstrate that materials obtained by
``site-substitution'' of t-PA, in which the hydrogen atoms of t-PA are replaced
with transverse conjugated groups will simultaneously have small optical
gaps {\it and} E(2A$_g$) $>$ E(1B$_u$). The physical reasoning for this is
as follows. Systems obtained by such site-substitution consist of units that
are molecular (for example,
trans-stilbene in the case of PDPA), and consecutive units are linked by
a single longitudinal bond. With moderate e-e interactions, the ground state
can be thought of as covalent (all atomic sites singly occupied). Optical
excitation involves intra- and inter-unit one-electron hoppings that
generate double occupancies on the units. Since the unit consists of a
large molecule in the case of the substituted polyene, this double occupancy
occupies the antibonding molecular orbital of the unit, which is delocalized
over the entire unit.
Thus the {\it effective}
on-site Coulomb interactions (the effective Hubbard interaction, hereafter
$U_{eff}$) is 
smaller,
and this can simultaneously give smaller E(1B$_u$) (relative to t-PA) and
E(2A$_g$) $>$ E(1B$_u$). This particular idea is related to a very
similar idea in the area of organic conducting
charge-transfer solids, where it has
long been believed that by going to a larger organic molecule [for example
from TMTTF, tetrathiafulvalene to BEDT-TTF, bis(ethylenedithio)fulvalene] 
the molecular Hubbard $U$ decreases \cite{Yamaji}.
The only difference in the present case is that we are comparing the
bare Hubbard $U$ (in t-PA) and a molecular $U_{eff}$ (in the substituted
material), and that the interunit one-electron hopping
here is much larger (comparable
to the intraunit hoppings) than in the charge-transfer solids, so that
molecular exciton ideas have to applied with caution.

\begin{figure}
\begin{center}
\epsfysize=4cm\epsfbox{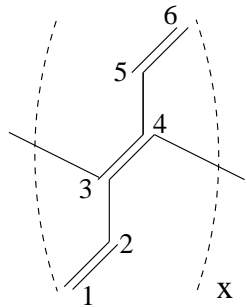}
\caption{The model substituted polymer, poly-diethylenepolyacetylene. The
numbers are indices for the atomic sites wthin each unit (see text).}  
\label{polymer}
\end{center}
\end{figure}

We consider here
the simplest possible $\pi$-conjugated polymer with conjugated sidegroup, the
hypothetical system poly-diethylenepolyacetylene, shown in Fig.~\ref{polymer},
in which all hydrogen atoms of t-PA have been replaced by ethylene groups.
The reason for choosing this hypothetical system is that precisely because
of its simple electronic structure, it can be considered
as the prototype
model for this class of (future) materials. In addition, in order
to illustrate
the theoretical result we are after
we need a physical system for which, (a)
accurate many-body computations can be done within a minimal basis (since for
any realistic system with large molecular units full configuration
interaction, FCI, with the complete basis is out of the question) for
relatively
long chains, and (b) proper justifications can be given for ignoring
basis functions
beyond the minimal basis. Note that (b) is at least as important as (a), since
it is known that the 2A$_g$ can occur above the 1B$_u$ also because of
incomplete CI \cite{Hudson,Srinivasan}.
Thus for our study to be complete it is essential that we prove
that our result is {\it not} a consequence of our choosing the minimal basis.
The symmetry and simplicity of our model system, along with the particular
basis we have chosen allow
us to achieve this.

In the next section we give the Hamiltonian, parameters and the scope of our
calculations. In section ~\ref{exciton} we discuss the complete
exciton basis for
oligomers of the system shown in Fig. ~\ref{polymer}, following which we
discuss our minimal basis, which in this case consists of the highest
occupied molecular orbital (HOMO) and lowest unoccupied molecular orbital
(LUMO) of each unit. In section \ref{numerics} we present our numerical
calculations for the unsubstituted and substituted polyenes to show that there
exists a broad range of Hubbard $U$ over which the 2A$_g$ is below the
1B$_u$ in the unsubstituted polyene, but above it in the substituted material
of Fig.~\ref{polymer}.
Wavefunction analysis within the exciton basis valence bond (VB) method
\cite{Chandross}
also shows that the effective correlations in the substituted polyene are
smaller. Following this, in section \ref{higher} we give very detailed
heuristic but rigorous arguments which show that this result is not a
consequence of incomplete CI. The mechanism of the 2A$_g$ -- 1B$_u$
crossover in polyenes is by now understood; specifically, the correlated 2A$_g$ state
is dominated by a particular {\it class} of two electron -- two hole (2e--2h)
excited configurations, viz., covalent configurations that do not contain any
double occupancy. Since the 1B$_u$
consists of ionic configurations only, there must exist a $U$ where the
1B$_u$ energy is higher. Because of the
pictorial natures of the wavefunctions
in the diagrammatic exciton basis VB method \cite{Chandross},
it is possible to classify
precisely all possible one electron -- one hole (1e--1h) and 2e--2h exciton
diagrams. As we show in section ~\ref{higher}, based on the symmetry
properties of our simple model system this information can then
be used to demonstrate that the higher energy MOs of the units
play insignificant roles
in the 2A$_g$ -- 1B$_u$ crossover and that the minimal basis results are
therefore accurate.

\section{The theoretical model}
\label{model}

We will consider our model polymer system within the simple dimerized Hubbard
Hamiltonian,
\begin{subequations}
\begin{eqnarray}
H = H_{1e} + H_{ee} \\
H_{1e} = -\sum_{\langle ij \rangle,\sigma}t_{ij}c_{i,\sigma}^\dagger c_{j,\sigma} \\
H_{ee} = U\sum_{i}n_{i,\uparrow}n_{i,\downarrow}
\label{Hubbard}
\end{eqnarray}
\end{subequations}
\noindent In the above, $H_{1e}$ describes one-electron nearest neighbor
hoppings of electrons, and $H_{ee}$ consists of the electron-electron
(e-e) interaction
within the Hubbard approximation. The Fermion operator
$c_{i,\sigma}^\dagger$ creates a $\pi$-electron
of spin $\sigma$ on site $i$, $n_{i,\sigma}$ is the number of electrons with
spin $\sigma$ on site $i$, and
$\langle .. \rangle$ implies nearest neighbors. The one-electron hopping
integrals
$t_{ij}$ are taken to be $t_1$ = 2.4(1 -- $\delta$) eV, and $t_2$ =
2.4(1 + $\delta$) eV, with $\delta$ = 0.07, corresponding to single and
double bonds, respectively. These values of the hopping integrals are
considered standard for $\pi$-conjugated systems.
We
have ignored all Coulomb interactions other than the on-site repulsion $U$,
since
the 2A$_g$ - 1B$_u$ crossover is related to this interaction only, with the
spin-independent long range intersite Coulomb interactions merely modifying the
magnitude of $U$ at which the crossover occurs. Nonzero intersite Coulomb
interactions make our exciton basis calculations much more complicated,
without providing fundamentally new insight.

Within the Hamiltonian of Eq.~(1), the 2A$_g$
occurs above the 1B$_u$ at $U$ = 0 in linear polyenes.
For nonzero $\delta$, as $U$ is increased from
zero, there exists a {\it finite} $U_c$ at which the 2A$_g$ - 1B$_u$ crossover
occurs, and the 2A$_g$ now occurs below the 1B$_u$ \cite{Soos1,Soos2,Shuai}. We 
perform
similar calculations for oligomers of the
the substituted materials, polydiethylenepolyacetylene (Fig.~\ref{polymer})
here. 
For our hypotheses in section \ref{intro}
to be correct the following should be true. First, not only should the optical
gap in the substituted polyene be smaller than that of the corresponding
polyene with the same number of backbone chain carbon atoms at all $U$, but
the rate of increase of E(1B$_u$) with $U$ must also be smaller. The latter
ia a definite signature of smaller effective correlation $U_{eff}$. Second,
if $U_{eff}$ is indeed smaller than the bare $U$ in the substituted polyene,
the bare $U_c$ at which the 2A$_g$ - 1B$_u$ crossover occurs in the substituted
polyene must be larger than in the unsubstituted polyene. As a consequence, for
a given e-e interaction it becomes possible to have the 2A$_g$ above the
1B$_u$ in the substituted polyene even as it is below the 1B$_u$ in the 
unsubstituted polyene. 

\section{Diagrammatic exciton basis for polydiethylene-acetylene}
\label{exciton}

\subsection{The complete basis}

It is clear that accurate many-body calculations for
oligomers with the structure of Fig.~\ref{polymer}  as the repeat unit
can only be done within a reduced basis. The reduced basis cannot be within the
configuration space, as it would be simply incorrect to ignore atomic orbitals.
A common approach in such cases is often to perform CI calculations using
a limited basis of molecular orbitals (MOs), but this is not suitable in the
present case: MOs are delocalized over the entire space, and within the
MO approach there is simply no information on the connectivity between the
atoms that can allow choosing the reduced basis. It is for this reason that
we have chosen the exciton basis VB method, which was previously used to perform
exact calculations within the Pariser-Parr-Pople model \cite{PPP1,PPP2}
for linear polyenes and
to obtain diagrammatic wavefunctions \cite{Chandross}. 
As noted there, the exciton
basis VB method is a hybrid of configuration space VB and the momentum-space MO
pictures, and thus the basis functions within this approach retain the full
information on the connectivity between the atoms (as in configuration space)
while at the same time, as in MO theory, the basis functions can be given certain
hierarchy from energetic considerations. As we show later, both of these are
essential for choosing the minimal basis as well as its justification.

As in the case of polyenes \cite{Chandross} we arrive at the exciton basis by
rewriting the one-electron term $H_{1e}$ in Eq. 1(b) as a sum of two 
terms, one intra-unit and the other inter-unit,
\begin{equation} 
H_{1e} = H_{1e}^{intra} + H_{1e}^{inter}
\label{split}
\end{equation}
The intra-unit part of the Hamiltonian is the simple H\"uckel Hamiltonian for
the hexatriene units,
\begin{equation}
H_{1e}^{intra} = -\sum_{\mu}\sum_{j=1}^{5}t_{j,j+1}[c_{\mu,j,\sigma}^{\dagger}c_{\mu,j+1,\sigma} + h.c.]  
\label{intra}
\end{equation}
where $\mu$ is an index for the units and $j$ is an atom within the unit. The
inter-unit part of $H_{1e}$ can be similarly written as,
\begin{equation}
H_{1e}^{inter} = -t_1\sum_{\mu}[c_{\mu,4,\sigma}^{\dagger}c_{\mu+1,3,\sigma} + h.c.]
\label{inter}
\end{equation}
where we recognize that the bonds between the units involve the weaker hopping 
integral $t_1$ only, and only the atom 4 of the unit of the left unit and the
atom 3 of the right unit are bonded (see Fig.~\ref{polymer} for the numbering
of the atoms within each unit). 

The solutions to $H_{1e}^{intra}$ are the intra-unit local
H\"{u}ckel MOs, described by,

\begin{equation}
\psi_{\mu,k,\sigma}^{\dagger} = \sum_{j=1}^{6} A_{kj}c_{\mu,j,\sigma}^{\dagger}
\label{MOs}
\end{equation}

where $k$ = 1 -- 6, with $k$ = 3 and 4 corresponding to the
HOMO and LUMO, respectively. The reverse transformation,
\begin{equation}
c_{\mu,j,\sigma}^{\dagger} = \sum_{k=1}^{6} A_{jk}\psi_{\mu,k,\sigma}^{\dagger}
\label{reverse}
\end{equation}
\noindent where $[A_{jk}] = [A_{kj}]^{-1}$ gives the site-operators in terms of the local
MO operators, and can be used to obtain $H_{1e}^{inter}$ in terms of these,
\begin{equation}
H_{1e}^{inter} = -t_1\sum_{\mu} \sum_{k,k',\sigma}[A_{4k}A_{3k'}\psi_{\mu,k,\sigma}^{\dagger}\psi_{\mu+1,k',\sigma} + h.c.]
\label{inter_mo}
\end{equation}

\noindent $H_{1e}^{inter}$ when operating on any exciton basis VB
diagram leads to
interunit charge-transfer (CT), which may involve pairs of bonding MOs,
antibonding MOs, or bonding and antibonding MOs. The magnitude, and more
importantly the sign, of the CT matrix element depend on the $A_{jk}$.
In the case of linear polyenes,
CT involving the bonding MO of the unit on the left lead to positive
matrix elements of $H_{1e}^{inter}$, whereas CT involving the antibonding MO
of the left unit lead to negative matrix elements \cite{Chandross}. The same
is true here if $k, k'$ in Eq.~(\ref{inter_mo}) are limited to the HOMO and LUMO
($k, k'$ = 3, 4).

Eq.~(\ref{reverse}) also enables us to rewrite the Hubbard interaction as,
\begin{widetext}
\begin{eqnarray}
H_{ee} = U \sum_{\mu} \sum_{j=1}^{6}[\sum_{k_1,k_2} A_{jk_1} A_{jk_2} \psi_{\mu,k_1,\uparrow}^{\dagger} \psi_{\mu,k_2,\uparrow}
\sum_{k_3,k_4} A_{jk_3} A_{jk_4} \psi_{\mu,k_3,\downarrow}^{\dagger} \psi_{\mu,k_4,\downarrow}] \nonumber\\
= U \sum_{\mu} \sum_j[\sum_{k_1,k_2}|A_{j,k_1}|^2|A_{j,k_2}|^2N_{\mu,k_1,\uparrow}N_{\mu,k_2,\downarrow}
+ \sum_{k,k_1,k_2}|A_{j,k}|^2 \cdot A_{j,k_1}A_{j,k_2}N_{\mu,k,\uparrow}\psi_{\mu,k_1,\downarrow}^{\dagger}\psi_{\mu,k_2,\downarrow}+\nonumber\\
+ \sum_{k,k_1,k_2}|A_{j,k}|^2 \cdot A_{j,k_1}A_{j,k_2}N_{\mu,k,\downarrow}\psi_{\mu,k_1,\uparrow}^{\dagger}\psi_{\mu,k_2,\uparrow}
+\sum_{k_1 \neq k_2}\sum_{k_3 \neq k_4} A_{j,k_1}A_{j,k_2}A_{j,k_3}A_{j,k_4}
\psi_{\mu,k_1,\uparrow}^{\dagger}\psi_{\mu,k_2,\uparrow}\psi_{\mu,k_3,\downarrow}^{\dagger}\psi_{\mu,k_4,\downarrow}]
\label{H_U}
\end{eqnarray}
\end{widetext}

Here $N_{\mu,k,\sigma} = \psi_{\mu,k,\sigma}^{\dagger}\psi_{\mu,k,\sigma}$, the
number of electrons with spin $\sigma$ in the $k$th MO of the $\mu$th unit.
Thus $H_{ee}$ contains diagonal density-density terms, as well as
off-diagonal density-electron transfer and
electron transfer-electron transfer terms within the exciton basis. All terms,
however, are local (i.e., intraunit), and electron-transfers due to the
Hubbard interaction are strictly between
the MOs of the same unit. If the units possess both center of inversion and
charge-conjugation symmetry (as is true here), an additional simplification
occurs when calculations are carried out within a reduced basis containing
pairs of MOs related by charge-conjugation symmetry \cite{Chandross}, viz.,
terms containing products of density and electron transfer vanish exactly
\cite{Chandross}. This is because the product $A_{j,k_1}A_{j,k_2}$ in the
density - electron transfer term
has different signs for sites $j$ that are related
by spatial symmetry. Of course when
considering the effects of the MOs not included in the reduced basis the
effects of these terms have to be considered.

\subsection{The minimal basis}

\begin{figure}
\begin{center}
\epsfysize=4cm\epsfbox{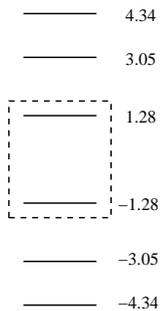}
\caption{The single-particle energies (in eV) for each isolated unit,
with standard hopping integrals (see text). The dashed box indicates our
minimal basis.}
\label{orbitals}
\end{center}
\end{figure}

The complete Hamiltonian, $H_{1e} + H_{ee}$ is way beyond our reach for any
system containing more that two units of the substituted polyene of 
Fig.~\ref{polymer}. At the same time, we show in section~\ref{higher}
that the two-unit oligomer is a special case, and is not of interest.
We therefore work within a minimal basis which is chosen as follows.
We first solve
$H_{1e}^{intra}$ and
note the following: (i) the HOMO-LUMO gap is much smaller
than the other bonding to antibonding
energy gaps (see Fig.~\ref{orbitals} for the single-particle
energies), and (ii) $H_{1e}^{inter}$ involves only atoms 3 and 4 of each unit,
and the contributions by the HOMO -- 1 and LUMO + 1 levels to the electron
densities on these atoms ($A_{j,k}$ in Eq.~(\ref{reverse}) for k = 2 and 5)
are relatively small [$|A_{j,1}| = |A_{j,6}|$ = 0.52, $|A_{j,2}| = |A_{j,5}|$ =
0.21, $|A_{j,3}| = |A_{j,4}|$ = 0.43, where $j$ = 3 and 4], although the
contributions by the outermost single-particle levels $k$ = 1 and 6
are large again (note,
however, that the single-particle energy gaps between these last pair of MOs
is rather large, from Fig.~\ref{orbitals}). We therefore construct our minimal
basis using only the HOMO and the LUMO
and retain only $k, k'$ = 3 and 4 in Eqs.~\ref{inter_mo} and ~\ref{H_U}.
Since these MOs are related by charge-conjugation symmetry the
density - electron transfer terms (but not the electron transfer -- electron
transfer term) in $H_{ee}$ vanish exactly within our minimal
basis and the minimal exciton basis Hamiltonian is very similar to the one
considered before for polyenes \cite{Chandross}, except that the $A_{jk}$
are very different now ($|A_{jk}|$ = 2$^{-1/2}$ in polyenes).
Note in particular that with $H_{ee}$ limited to the Hubbard interaction the
remaining electron transfer -- electron transfer term in $H_{ee}$ consists
of two electron transfers involving opposite spins only, and therefore 
virtual excitations involving the outer MOs do not occur.
The lower occupied MOs therefore contribute a constant to the
diagonal matrix elements of the exciton basis VB diagrams, which is the
frozen core approximation.

While single-particle energies and electron densities give
guidance to the choice of the minimal basis,
{\it justification} of the minimal basis requires the demonstration
that the minimal
basis can satisfactorily reproduce the results expected for the complete
Hilbert space. This is easily done for the one-electron part of the
Hubbard Hamiltonian, $H_{1e}$.
We solve the full H\"uckel Hamiltonian for oligomers with number of
backbone carbon atoms N = 4 -- 10, and compare the exact
lowest excitation energies with the approximate excitation energies
obtained using the minimal exciton basis. Excellent agreement is found, as is
shown in Table 1, where we compare the energies of
1e--1h B$_u$ excitations for N = 4 -- 10 (the 1e--1h
A$_g$ excitation energies can be obtained from the B$_u$ energies and are
therefore not given separately).

\begin{table*}
\begin{center}
\begin{tabular}[c]{|c|c|c|c|c|c|c|c|c|}\hline
States  & \multicolumn{2}{c|}{N = 4} & \multicolumn{2}{c|}{N = 6} & \multicolumn{2}{c|}{N = 8} & \multicolumn{2}{c|}{N = 10}\\ \cline{2-9} & Exact H\"{u}ckel
& Exciton VB  & Exact H\"{u}ckel & Exciton VB & Exact H\"{u}ckel & Exciton VB & Exact H\"{u}ckel & Exciton VB\\\hline
$1B_u$ & 1.81 & 1.86 & 1.41 & 1.54 & 1.17 & 1.36 & 1.02 & 1.25\\
$2B_u$ & 3.37 & 3.53 & 2.48 & 2.69 & 2.22 & 2.35 & 1.99 & 2.06\\
$3B_u$ &      &      & 2.89 & 2.83 & 2.49 & 2.37 & 2.16 & 2.08\\
$4B_u$ &      &      & 3.55 & 3.72 & 3.05 & 3.18 & 2.33 & 2.66\\
$5B_u$ &      &      &      &      & 3.27 & 3.22 & 2.81 & 2.84\\
$6B_u$ &      &      &      &      & 3.62 & 3.34 & 2.97 & 2.91\\ \hline
\end{tabular}
\end{center}
\caption{Exact versus minimal exciton basis
energies (in eV) of the lowest 1e-1h B$_u$ states in oligomers of the substituted
polyene at U = 0.
}
\end{table*}
Justification of
the minimal basis even when $H_{ee}$ is nonzero is much more complicated. We
postpone this until after we present our numerical demonstration
within the minimal
basis that the 1B$_u$ optical
gap is smaller in the substituted polyene even when $U$ is included, and the
$U_c$ at which the 2A$_g$ becomes lower than the 1B$_u$ is much larger for the
substituted system. As mentioned in section \ref{intro}, we are able to give a
very complete proof that our reasult of higher $U_c$ is not a consequence
of incomplete CI, based on symmetry properties of exciton basis VB diagrams.

\subsection{The mechanism of the 2A$_g$ -- 1B$_u$ crossover}

Although in reference \onlinecite{Chandross} detailed
discussions of exciton basis VB diagrams for the case of linear polyenes
were presented, in view of what follows in this and the next sections, it is
useful to reemphasize some of these. We therefore discuss
the mechanism of the 2A$_g$ -- 1B$_u$ crossover within the the
exciton basis VB approach. While
it is known that the occurrence of the 2A$_g$ below the 1B$_u$ is a
correlation effect that introduces strong configuration mixing between the
``ground state'' configuration and 2e--2h configurations, a key theme here is
that only a specific {\it class} of 2e--2h states is actually relevant.

Polyene exciton basis VB diagrams, and the substituted polyene exciton basis
VB diagrams within the minimal basis are coupled by $H_{1e}^{inter}$ and
$H_{ee}$.
We give here the complete expressions for $c_{\mu,3,\sigma}$ and
$c_{\mu,4,\sigma}$ such that the effects of $H_{1e}^{inter}$ (see
Eq.~(\ref{inter_mo})) can be understood.

\begin{subequations}
\begin{eqnarray}
c_{\mu,3,\sigma} = -0.52\psi_{\mu,1,\sigma} + 0.21\psi_{\mu,2,\sigma}
+ 0.43\psi_{\mu,3,\sigma}\nonumber\\
+ 0.43\psi_{\mu,4,\sigma} - 0.21\psi_{\mu,5,\sigma}
-0.52\psi_{\mu,6,\sigma}\\
c_{\mu,4,\sigma} = -0.52\psi_{\mu,1,\sigma} - 0.21\psi_{\mu,2,\sigma}
+ 0.43\psi_{\mu,3,\sigma}\nonumber\\
- 0.43\psi_{\mu,4,\sigma} - 0.21\psi_{\mu,5,\sigma}
+ 0.52\psi_{\mu,6,\sigma}
\end{eqnarray}\label{operators}
\end{subequations}

All numerical coefficients in the above are obtained from the invert of
the H\"uckel
wavefunction matrix for linear hexatriene. 
The relative signs (though not magnitudes)
between the coefficients of the
bonding and antibonding MOs in the site operators for the polyenes are the
same as those between the coefficients of the HOMO and LUMO
($\psi_{\mu,3,\sigma}$ and $\psi_{\mu,4,\sigma}$) for the substituted
polyene in the above. Thus in the following both systems can be described
simultaneously. 
Note that $H_{1e}^{inter}$ from Eqs.~(\ref{inter_mo}) and (9)
consists of terms $\psi_{\mu,k,\sigma}^{\dagger}\psi_{\mu+1,k',\sigma}$, which
can be both positive or negative. These signs
will be important in what follows in
this section and in section \ref{higher}.

\begin{figure}
\begin{center}
\epsfysize=4cm\epsfbox{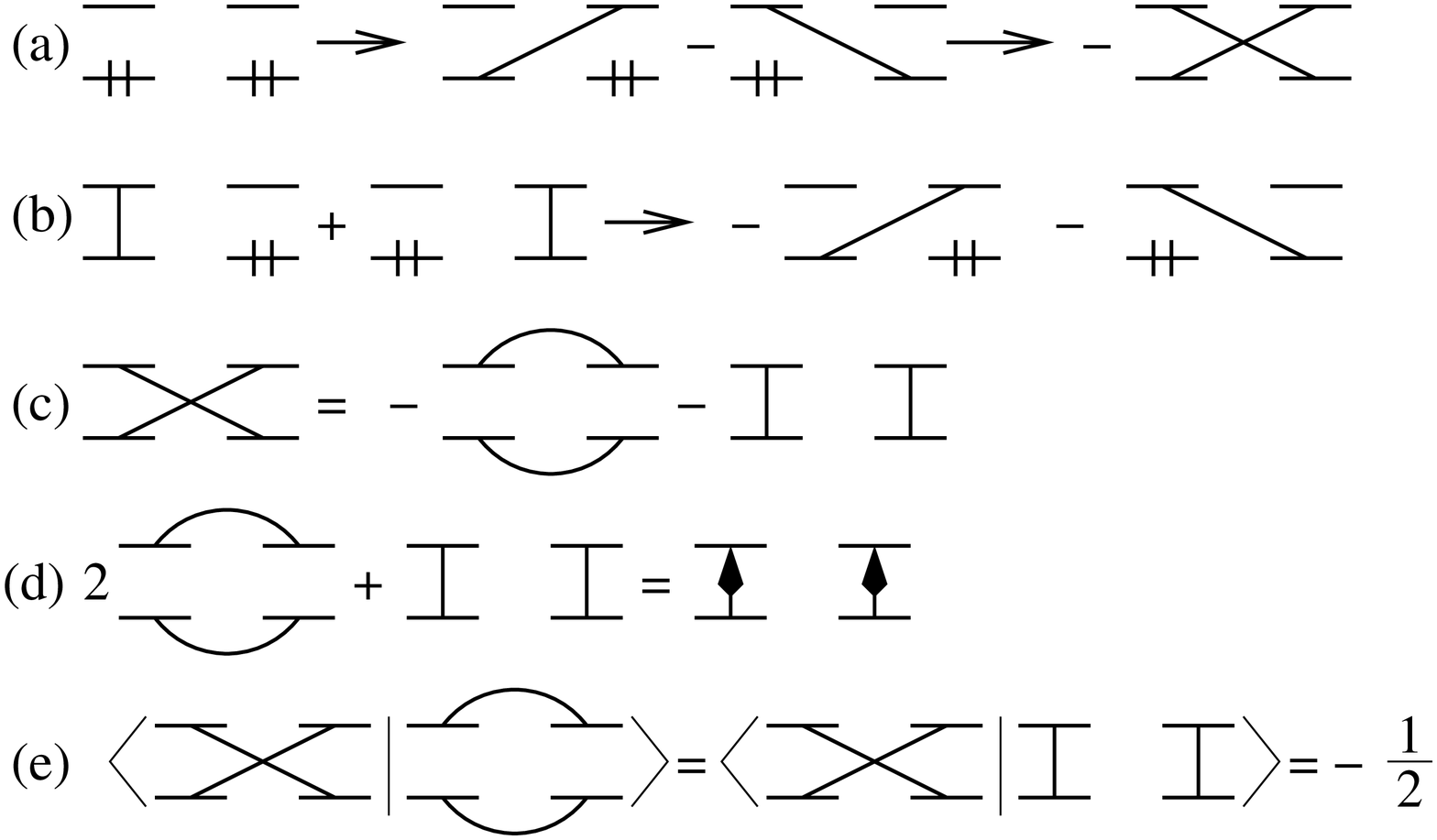}
\caption{CI channels and exciton basis VB diagrams relevant for the
2A$_g$ -- 1B$_u$ crossover in a two-unit system with two MOs per unit. Bonding
and antibonding MOs of each unit are occupied by 0, 1 and 2 electrons, with
pairs of singly occupied sites coupled by singlet bonds (see text). Each arrow
in (a) and (b) represents one application of $H_{1e}^{inter}$ on the diagrams
on the left, to obtain the diagrams on the right of the arrow. The plus
and minus signs of the matrix elements are obtained from Eq.~(9),
and their magnitudes are not shown. The linear relationship (c) shows that
the diagram with crossed bonds is a superposition of two other diagrams with
the same orbital occupancies but with noncrossing bonds. The diagram on the
right in (d) is triplet--triplet (TT), with two triplets localized on the two
units. (e) shows the overlap between the nonorthogonal 2e--2h VB diagrams.}
\label{basis}
\end{center}
\end{figure}

In Fig.~\ref{basis} we show the VB exciton diagrams
and CI channels that are most relevant for the 2A$_g$ -- 1B$_u$
crossover in a
two-unit system, which can be either the two-unit polyene or the two-unit
substituted polyene with the lower levels completely filled.
Each arrow in this and similar figures below corresponds
to one
application of $H_{1e}^{inter}$ (Eq.~(\ref{inter_mo})) on the VB diagrams 
to the immediate
left, to get the VB diagramis to the immediate right of the arrow.
As mentioned above, the plus and minus signs here are obtained fron the
equations for the site-operators above.
Aside from the signs the CI channels should also include the
magnitudes $A_{4k}A_{3k'}$, which can again be obtained from
Eq.~(\ref{operators}), but which are not very important for our discussions.

The first diagram in CI channel (a) is the product
wavefunction of the ground
states of two noninteracting units, with all electrons in the bonding MOs,
which we refer to hereafter as the
Simpson ground state \cite{Simpson}. This diagram is symmetric with
respect to the mirror plane between the two units, occurs only in the
$A_g$ subspace, and dominates the 1A$_g$ wavefunction for small to moderate
$U$. One application of $H_{1e}^{inter}$ generates the ``minus'' linear
combination of two charge-transfer (CT) diagrams (two because CT can occur
in either direction). The relative minus sign is obtained from
Eq.~(9).
Here and in the following
a singlet bond, drawn as a line connecting MOs $k$ and $k'$, 
on the same or different
units, is defined as
2$^{-1/2}(\psi_{\mu,k,\uparrow}^{\dagger}\psi_{\mu',k',\downarrow}^{\dagger} -
\psi_{\mu,k,\downarrow}^{\dagger}\psi_{\mu',k',\uparrow}^{\dagger})|0\rangle$
as before \cite{Chandross}.
A second application of $H_{1e}^{inter}$, such that CT is now in the
{\it opposite} direction, generates the 2e--2h
charge-neutral diagram with crossed
bonds (the last diagram in (a)) which can again occur in the A$_g$ space
only. The ``minus'' sign
in front of this diagram in the CI channel shown occurs in the case
of both unsubstituted and substituted polyenes, and will be relevant for
later discussions.

Fig.~\ref{basis}(b) shows CI channels originating from the superposition of
two Frenkel
exciton diagrams with intra-unit excitations.
The two Frenkel exciton diagrams
are absent in the A$_g$ subspace, and the
``plus'' linear combination shown occurs in the B$_u$ susbspace.
Operating with $H_{1e}^{inter}$ on the two Frenkel exciton diagrams generates
the same CT diagrams that are generated in the second step in (a), but
now with a relative plus sign between them. Because of this relative plus sign
a second application of
$H_{1e}^{inter}$ (again promoting CT in the opposite direction from the first
step) leads to an exact cancellation, such that the diagram with crossed
bonds does not occur in the B$_u$ subspace.

The mechanism of the 2A$_g$ -- 1B$_u$ crossover is then as follows. 
The CT diagrams in Fig.~\ref{basis} have unequal numbers of electrons on the
units, and are hence necessarily ionic in the language of configuration
space VB theory. The Frenkel exciton diagrams are also ionic, as can be
confirmed by direct expansion of exciton basis operators \cite{Chandross}.
The diagram with the crossed bonds in Fig.~\ref{basis}(a), 
however, is charge-neutral as far as the
units are concerned, and as shown in Fig.~\ref{basis}(c) is a superposition
of two VB diagrams with noncrossing bonds, which can also be confirmed
by expansion of exciton basis operators. Equally importantly, a 2:1 
superposition of these last two diagrams (see Fig.~\ref{basis}(d))
corresponds to two triplets, or
triplet-triplet (TT) excitation on the two units \cite{Chandross}, which is
covalent in the language of configuration space VB theory. As has been
emphasized by previous authors, the
2A$_g$ -- 1B$_u$ crossover can then be understood as a  
competition between CT and TT diagrams. Even though the diagram with crossed
bonds is obtained in second order in $H_{1e}^{inter}$ from the
Simpson ground state, this second order process can be more important than 
the first order process leading to the CT diagrams, because of relationship
(d), since covalent diagrams have smaller diagonal matrix element of $H_{ee}$
than ionic diagrams. The 2A$_g$ is dominated by ionic diagrams at small
$U$ and occurs above the 1B$_u$, while it is dominated by covalent TT
diagrams at moderate to large $U$ and occurs below the 1B$_u$.
We have discussed the crossover
for a two-unit system only. The only new feature in long chains
is the occurrence of CT and TT diagrams with long bonds between units that
are not neighbors (see reference \onlinecite{Chandross} and below).

\begin{figure}
\begin{center}
\begin{tabular}{l}
\epsfysize=4cm\epsfbox{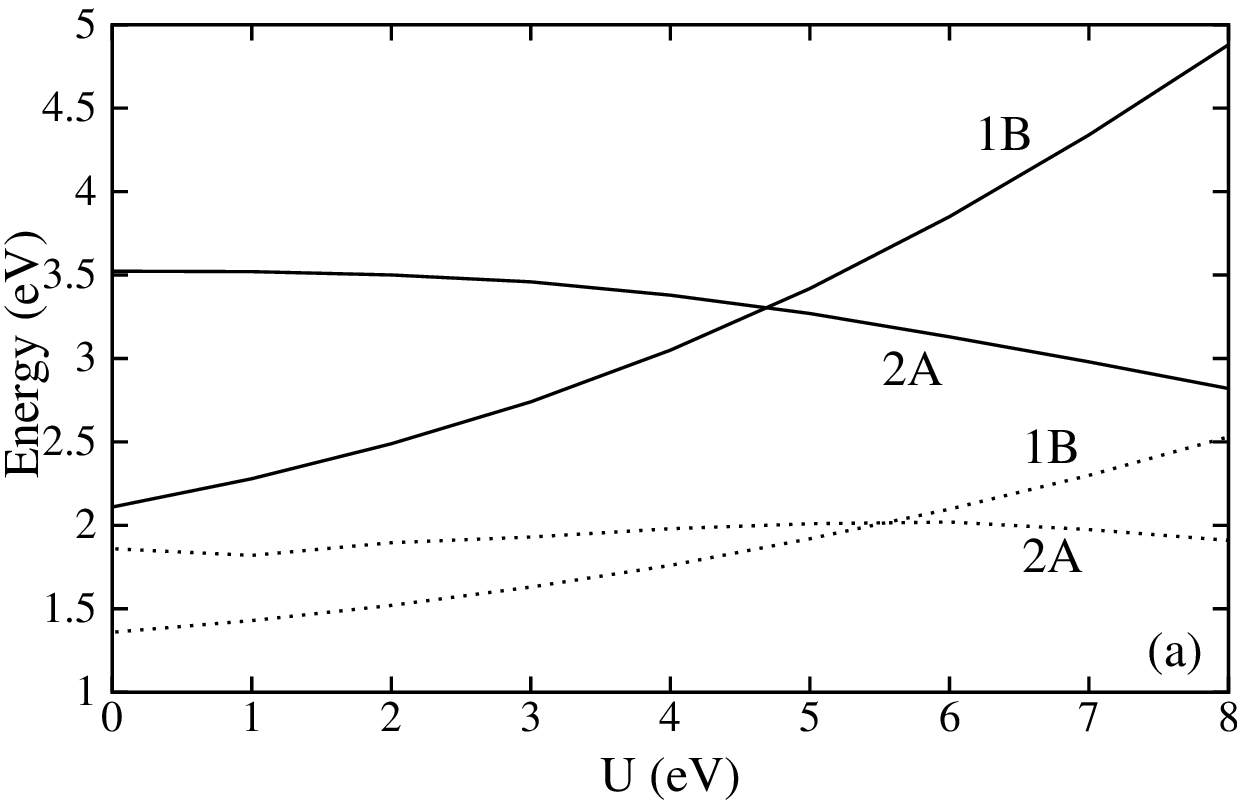} \\
\epsfysize=4cm\epsfbox{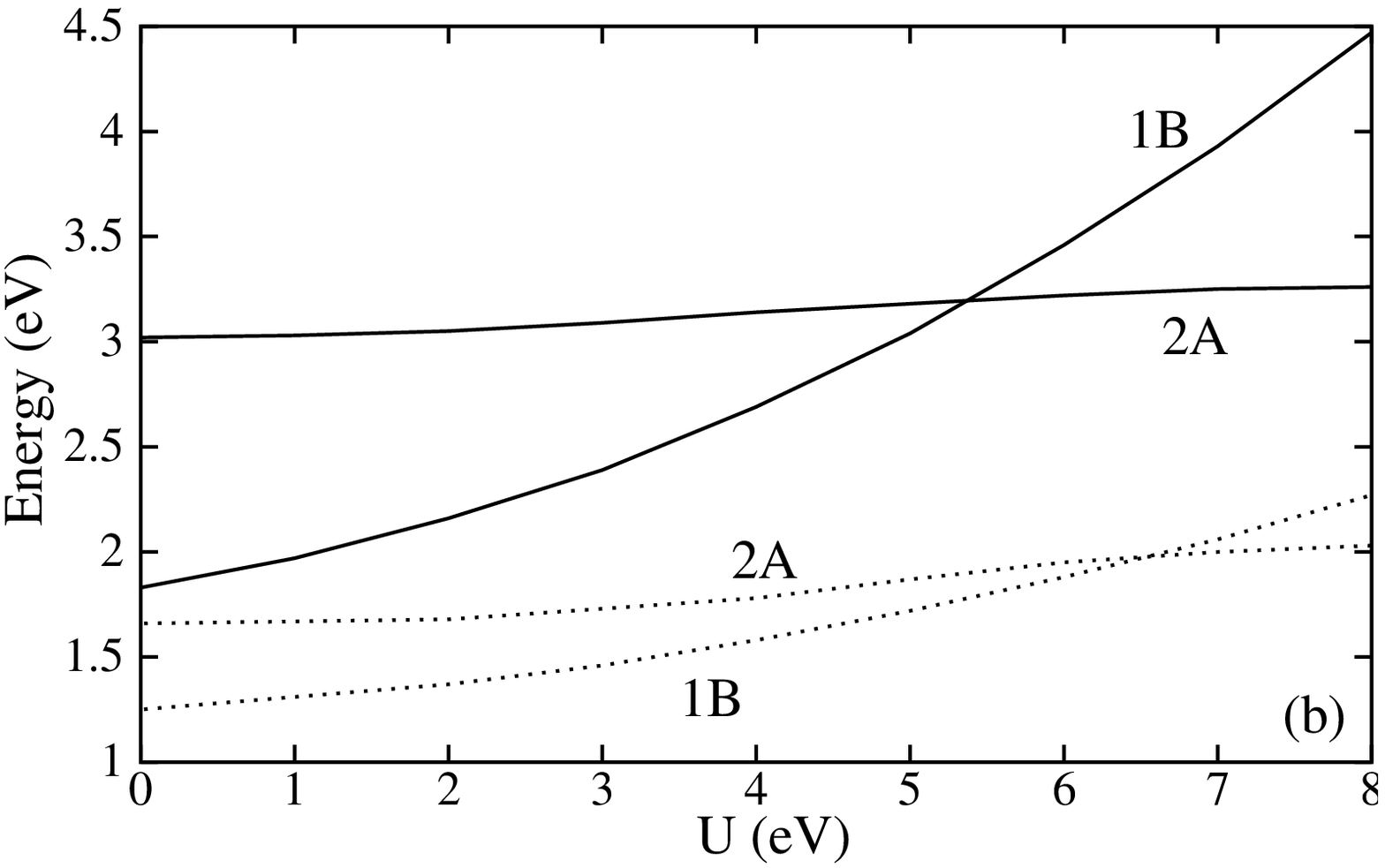}
\end{tabular}
\caption{The excitation energies E(1B$_u$) and E(2A$_g$) for
the unsubstituted (solid lines) and the substituted polyene (dotted lines), (a) N = 8, (b) N = 10.}
\label{energies}
\end{center}
\end{figure}

\section{Numerical results within the minimal basis}
\label{numerics}

\begin{figure}
\begin{center}
\epsfysize=4cm\epsfbox{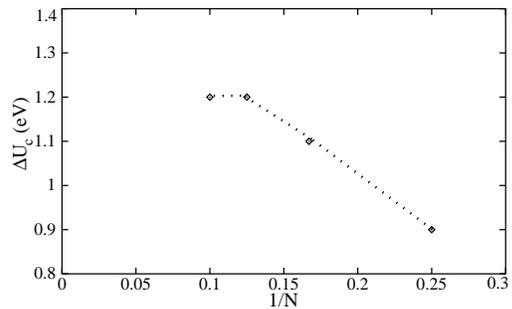}
\caption{The difference in $U_c$ between substituted and unsubstituted
polyene as a function of $1/N$.}
\label{DeltaU_c}
\end{center}
\end{figure}

We performed exact numerical calculations for N = 4, 6, 8 polyenes and
quadruple-CI (QCI) calculations for the N = 10 polyene to determine $U_c$.
Within the exciton
basis the Hamiltonian matrix is nonsparse and this is the reason for doing
QCI instead of exact calculations for N = 10. As has been shown before
\cite{Tavan,Guo}, QCI is excellent for determining the energies of
the lowest excitations, the 1B$_u$ and the 2A$_g$, to high accuracy,
at these N.
Identical calculations
were done within the minimal basis for oligomers of
polydiethylene-acetylene. In
Fig.~\ref{energies}(a) and (b)
we have plotted the excitation energies E(1B$_u$) and
E(2A$_g$), with respect to the ground state energy (E(1A$_g$) = 0) for
N = 8 and 10, respectively. As claimed in section ~\ref{intro}, E(1B$_u$) is
lower for the
substituted polyene in both N = 8 and 10, for all values of $U$. Note in
particular that E(1B$_u$) for the substituted polyene increases with the
bare $U$ much
less rapidly than for the unsubstituted polyene, which is a distinct signature
that $U_{eff}$ in the substituted material is smaller. The $U_c$ at which
E(2A$_g$) becomes smaller than E(1B$_u$) in the unsubstituted polyenes is
very slightly larger in N = 10 than in N = 8, in agreement with previous
work \cite{Shuai}. 
Importantly, the 2A$_g$ remains above
the 1B$_u$ in the substituted polyene for a much
larger range of the Hubbard $U$,
and $U_c$ for the substituted polyene is larger
than that of the unsubstituted polyene by about 1 eV. The initial increase
in E(2A$_g$) with $U$ for substituted polyenes
in Fig.~\ref{energies} is real and has been seen 
previously in density matrix renormalization group (DMRG) studies of long
polyene chains by Shuai et. al. \cite{Shuai} (see, in particular,
Fig.~1 in this reference). As discussed there, this initial
increase is related to the magnitude of the 2A$_g$ -- 1B$_u$ gap at $U$ = 0.
The smaller this energy gap, the more the tendency to initial increase in 
E(2A$_g$)
(although E(2A$_g$) does begin to decrease again at large $U$). The
substituted polyenes considered here have relatively
small 2A$_g$ -- 1B$_u$ gap already
at N = 10, which explains the behavior of E(2A$_g$) in Fig.~\ref{energies}.
Taken together, the initial increase of E(2A$_g$), and the less rapid rise 
in E(1B$_u$) give the larger $U_c$ in the substituted polyenes. 
In Fig.~\ref{DeltaU_c} we have
plotted the difference $\Delta U_c$ between the $U_c$ values for the
substituted and the unsubstituted polyenes against the chain length N.
$\Delta U_c$ increases modestly between N = 4 and N = 8, but then is seen to
have a tendency to saturation. This behavior of $\Delta U_c$
is to be expected from what we have said in the above.
The initial increase in $\Delta U_c$ is due largely to the initial increase
in E(2A$_g$) with $U$ in the substituted polyene, which occurs for a larger 
range of $U$ the longer the system is. This in turn is simply a consequence
of smaller one-electron 2A$_g$ -- 1B$_u$ gap as N increases.
As the chain gets longer, the
rate of decrease of the one-electron 2A$_g$ -- 1B$_u$ gap itself decreases
and the range
of $U$ over which E(2A$_g$) has initial increase in the
substituted polyene tends to saturate, leading to a tendency of saturation
in the $U_c$ as well as $\Delta U_c$.  
Within the minimal basis therefore, $\Delta U_c$ is at least as 
large as that at
N = 10 in the long chain limit. 
The larger $U_c$ in the substituted polyenes is a direct consequence of the
delocalized HOMO and LUMO in the monomer, an effect which is included in the
$A_{jk}$. 

Information on the effective Coulomb correlation can also be found from
wavefunction analysis.
In our previous work we had studied linear polyenes within the
Pariser-Parr-Pople model \cite{PPP1,PPP2}, for both
small bond-alternation ($\delta$ =
0.07, where the 2A$_g$ is below the 1B$_u$,
and large bond-alternation ($\delta$ = 0.3), for which the 2A$_g$ is above the
1B$_u$. We had found that while the difference in the 1B$_u$ wavefunctions in
the two cases was very small, the difference between the 2A$_g$ wavefunctions
was substantial: the 2A$_g$ for small $\delta$ is dominated by TT exciton
basis diagrams, while for large $\delta$ there occurs a strong admixing of 
1e--1h CT
diagrams. It is then expected that similar
differences in the 2A$_g$ wavefunctions
would occur here for the unsubstituted and substituted polyenes. In
Fig.~\ref{1Bu} we show the 1B$_u$ wavefunctions for the N = 10 polyene and the
corresponding substituted polyene for $U$ = 6 eV, which is larger than the
$U_c$ for the polyene but smaller than the $U_c$ for the substituted polyene.
These wavefunctions appear to be very similar in nature.
In contrast, the 2A$_g$ wavefunctions for the two systems for the same $U$,
are remarkably different, as seen in Fig.~\ref{2Ag}.
While the polyene 2A$_g$ is predominantly TT and TT $\otimes$ CT
(diagrams which appear to be similar to the TT but in which the lengths
of the two interunit bonds are unequal are referred to as
TT $\otimes$ CT \cite{Chandross}) 
there is much stronger admixing of CT diagrams in the case of the
substituted
polyene, indicating clearly that the effective e-e correlation
in the latter is smaller for the same bare $U$. CT diagrams are necessarily
ionic in the language of configuration space VB theory \cite{Chandross,Mukho},
and the greater ionicity of the 2A$_g$ wavefunction in the substituted polyene
is related to its initial rise in energy with $U$ \cite{Shuai}.

\section{Beyond the minimal basis, the role of higher excitations.}
\label{higher}
\begin{figure*}
\begin{center}
\input{Fig6.tex}
\caption{The 1B$_u$ wavefunctions for the N = 10 (a) polyene, and (b)
substituted polyene within the minimal basis calculation, for $U$ = 6 eV.
Notice that the wavefunctions are similar in nature.}
\label{1Bu}
\end{center}
\end{figure*}

\begin{figure*}
\begin{center}
\input{Fig7.tex}
\caption{The 2A$_g$ wavefunctions for the N = 10 (a) polyene, and (b) 
substituted polyene within the minimal basis calculation, for $U$ = 6 eV.
Pairs of diagrams within the square parentheses occur as nearly 2:1 linear
combinations, reflecting TT character. Note that the two wavefunctions are
now very different, with CT contribution in (b) much larger than in (a).}
\label{2Ag}
\end{center}
\end{figure*}

A large body of previous work has established that in the case of linear
polyenes low order CI using a MO basis
does not capture important correlation effects \cite{Hudson,Srinivasan}.
Thus within the
the PPP Hamiltonian, double-CI fails to find the 2A$_g$ below the 1B$_u$ for
polyenes with N $>$ 8, and either QCI or MRD-CI becomes necessary to obtain the
correct excited state ordering. Similarly,
full-CI calculations within the Hubbard model for N = 12
using the six innermost bonding and antibonding MOs also fails to find the
correct excited state ordering. \cite{Srinivasan} Thus
the very fact that the minimal exciton basis gives a 2A$_g$ -- 1B$_u$
crossover at all is already significant. Nevertheless, since for each N in
polydiethylene-acetylene we ignore more MOs than we retain,
it might appear that the numerical result that
$U_c$ for this system is much larger than that of polyenes is a simple
consequence of our approximation. We show in this section that this is not
so, and the higher exciton basis VB diagrams play an insignificant role
in the 2A$_g$ -- 1B$_u$ crossover, especially in long chains.
There is a fundamental difference between
the CI involving completely delocalized MOs and the CI involving the
exciton basis VB diagrams. The latter possess local character and
are more suitable for calculations involving local Coulomb interactions.
We emphasize that even though our discussions below are heuristic
and based on physical arguments rather than numerical calculations, they are
complete and rigorous.

Our demonstration of the weak role of the higher exciton basis VB diagrams
will be done in several steps. First, we look at simple polyenes more
closely, to determine that only a certain class of higher excitations can
lower E(2A$_g$) with respect to the 1B$_u$. Based on this we arrive at
two specific criteria for relevance of higher excitations in
polydiethylene-acetylene, viz., that they
should also involve crossed bonds (in other words, that the relevant
configurations are those that are related to the high energy TT diagrams);
and that they are coupled to the lower energy TT diagrams. This allows us to 
severely limit our search of relevant high energy configurations. We then go
through the {\it complete list} of such configurations, both intra- and
interunit, asymmetric and symmetric with respect to the mirror-plane symmetry.
Because the two-unit oligomer is a special case (see below) we discuss this
separately. Following this we discuss longer oligomers. Our basic conclusion
is that the longer the oligomer, the more irrelevant are the high energy 
configurations. The discussion that follows is unavoidably complex, and
we have therefore given a separate summary at the end of this section. The
reader uninterested in the details of CI theory and the exciton VB basis can
directly go to this summary.

\begin{figure}
\begin{center}
\epsfysize=1.8cm\epsfbox{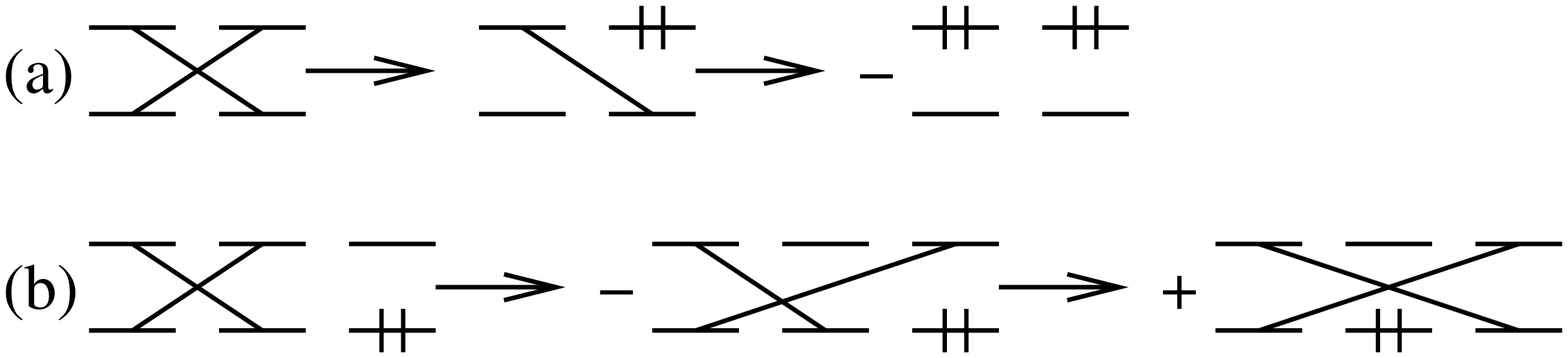}
\caption{CI channels that lower the energy of the 2A$_g$. Arrows have the
same meaning as in Fig.~\ref{basis}. (a) The CI process between the two-unit
symmetric 2e--2h excitation and the quadruple excitation. (b) Similar CI
process in the three-unit case that gives the symmetric 2e--2h diagram with
long bonds.}
\label{CI}
\end{center}
\end{figure}

Before we present our discussion of higher MOs in polydiethylene-acetylene,
we present additional discussions on the 2A$_g$ -- 1B$_u$ crossover in
polyenes that go beyond Fig.~\ref{basis}. Double-CI is not suitable for
correct evaluation of E(2A$_g$) since quadruple excitations that are coupled
to the doubly excited configurations, and that can therefore lower E(2A$_g$),
are excluded in double-CI. The CI channel describing this is shown in
Fig.~\ref{CI}(a), where the quadruply excited diagram for the
two-unit case is obtained from two applications of $H_{1e}^{inter}$
on the VB diagram with crossed bonds in Fig.~\ref{basis}, once again
such that the two CT processes are in the opposite direction. E(2A$_g$) is
also lowered by going to larger N. As shown in Fig.~\ref{CI}(b) this is due
to CI between the two diagrams with short nearest neighbor and long
next-nearest neighbor crossed bonds, again through two applications of
$H_{1e}^{inter}$ in opposite directions. Very similar CI process occurs between
the final diagram in Fig.~\ref{CI}(b) and the diagram with even longer
crossed bonds and so on.
Two things will be relevant in our
discussions below. First, even though diagrams with
long crossed bonds are not directly coupled to the Simpson ground state,
their relative weights 
(more precisely, the relative weights of the TT diagrams with long bonds)
in the exact 2A$_g$ can be large (see Fig. ~\ref{2Ag}).
This is because of the degeneracy of all 2e--2h VB diagrams with crossed
bonds within $H_{ee} + H_{1e}^{intra}$ which makes CI channels of the type in
Fig. \ref{CI}(b) extremely important. Second, the relative signs of the
diagram with nearest neighbor crossed bonds and the diagram with long crossed
bonds in Fig.~\ref{CI} are the same. The
same sign between them signifies constructive interference between the
CI channels in Fig.~\ref{basis}(a) and Fig.~\ref{CI}(b), and as seen
in Fig.~\ref{2Ag}, all TT diagrams have the same sign in the 2A$_g$
wavefunction (TT $\otimes$ CT diagrams, defined in the above
as TT-type diagrams that
are accompanied by CT, as is true for the pair within the first parentheses
in Fig.~\ref{2Ag}, can have opposite signs due to the additional CT),
while the relative sign of the Simpson ground state is opposite
(the relative sign of the Simpson ground state and the diagrams with
crossed bonds are the same in the 1A$_g$, as suggested by the CI channels;
see reference \onlinecite{Chandross}).

In order to understand the role of additional MOs in polydiethylene-acetylene
we will use the underlying information in Fig. \ref{CI}. Two classes of 2e--2h
excitations are
relevant for the 2A$_g$ -- 1B$_u$ crossover in the substituted
polyene, intra- and interunit.
The former are reached by application of $H_{ee}$
(all terms in Eq.~(\ref{Hubbard})) to the Simpson ground state,
the latter once again through two
applications of $H_{1e}^{inter}$. The intramolecular 2e--2h excitations that
are relevant for the 2A$_g$ -- 1B$_u$ crossover are known from the existing
literature on CI calculations using the MO basis for small molecules
\cite{Hudson}. As far as the interunit high energy 2e--2h excitations
are considered, the following is obvious from the previous sections and
reference \onlinecite{Chandross}:

{\it The only interunit 2e--2h VB diagrams that are relevant for the
2A$_g$ -- 1B$_u$
crossover are the ones with crossed bonds between units, since these are
the only interunit diagrams that are charge-neutral. Strong CI involving
diagrams with crossed bonds is essential for having contributions from higher
energy TT diagrams.}

Having identified the class of 2e--2h diagrams within the complete basis
that may be relevant for the 2A$_g$ -- 1B$_u$ crossover we note the following.
Unless the higher energy 2e--2h diagrams are directly coupled to the
lower energy 2e--2h diagrams by $H_{ee}$ or $H_{1e}^{inter}$ they have a very
weak effect on E(2A$_g$). In the absence of CI between the high and low energy
diagrams, we have a nearly block-diagonal Hamiltonian, with each block
containing its own set of excitations, with only the
matrix elements of the Simpson ground state linking the blocks. Thus the
2A$_g$ wavefunction continues to have strong CT contribution as shown in 
Fig.~\ref{1Bu}. Thus E(2A$_g$)
is lowered only if there exists CI channels analogous to those in Fig \ref{CI}
that couple the high and low energy VB diagrams with crossed bonds. This
leads to our second conclusion:

{\it Unless the high energy relevant 2e--2h diagrams, containing intra-
or interunit two-excitations, are coupled to the
low energy diagrams with crossed bonds through two applications of
$H_{1e}^{inter}$, they play an insignificant role in the 2A$_g$ -- 1B$_u$
crossover.}

We have limited ourselves only to second order processes between
excitations (fourth order from the
ground state) in the above, since
if the higher energy diagrams are coupled in even higher order, their effects
on E(2A$_g$) are even weaker than quadruple excitations involving the minimal
basis. Also, the couplings have to be necessarily through $H_{1e}^{inter}$,
since all couplings through $H_{ee}$ are intra-unit only.
The above restrictions allow us to severely restrict our search for
relevant high energy 2e--2h excitations and still reach the correct overall
conclusion concerning the role of MOs beyond the minimal basis. In the
following we discuss the case of the two-unit system first, followed by the
more general case of arbitrary chain length. Our overall conclusion is
that because of the
{\it local} nature of $H_{1e}^{inter}$ within the VB exciton basis, the
effect of the couplings with the high energy 2e--2h diagrams in long chains
is very weak. Because of the very detailed nature of the discussions that
follow, we have given a summary of these discussions in the last subsection.

\subsection{Two units: Coupling to intra-unit 2e--2h states}

\begin{figure}
\begin{center}
\epsfysize=7cm\epsfbox{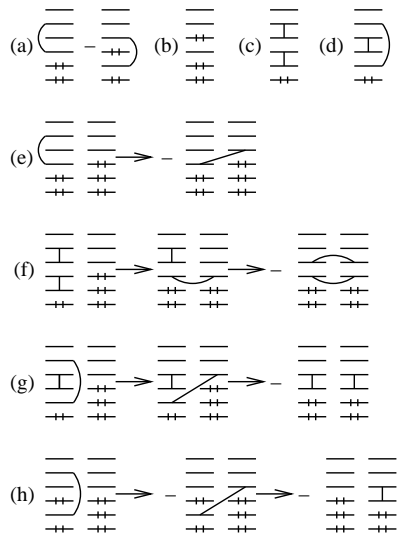}
\caption{(a) - (d): Intraunit excitations that dominate the molecular 2A$_g$
in hexatriene.
All 6 MOs are shown now. The lowest three in each unit are bonding MOs, while
the upper three are antibonding MOs.
(e): CI channel between one of the single excitations of (a) and a minimal
basis CT diagram, for the two-unit system.
(f) and (g): CI channels between intraunit 2e--2h excitations
and the two components of the TT diagram within the minimal basis, also for
the two-unit case. (h) CI channels between higher energy B$_u$ Frenkel
configuration and the minimal basis B$_u$ Frenkel configuration.
Arrows have the usual meaning in (e) -- (h).}
\label{channel1}
\end{center}
\end{figure}

Exact 2A$_g$ wavefunctions for small polyenes, obtained by FCI using the
MO basis have been
obtained by Schulten et. al. \cite{Hudson}.
The dominant components of the 2A$_g$
in molecular hexatriene are shown in Fig.~\ref{channel1}(a) -- (d). 
Of these, (a) is a
superposition of 1e--1h excitations
that can only {\it increase} E(2A$_g$).
As shown in Fig.~\ref{channel1}(e)
this 1e--1h configuration is coupled to intermolecular CT states in first
order by $H_{1e}^{inter}$.
The 2e--2h configuration (b) has been retained
in our minimal basis calculation. This leaves the molecular configurations
(c) and (d) in Fig.~\ref{channel1}, and as shown in Fig.~\ref{channel1} (f) and
(g), these are coupled to components of the TT diagrams in the minimal basis
by $H_{1e}^{inter}$
in second order. Thus both 1e--1h and 2e--2h higher energy molecular A$_g$
excitations are coupled to the minimal basis A$_g$ diagrams, and
without actual calculation, it is difficult to say which of
the two CI processes involving the molecular A$_g$ states
shown in Fig.~\ref{channel1} will dominate near $U = U_c$. Furthermore, there
occurs also CI between higher molecular B$_u$ configurations
and the minimal basis Frenkel configurations, as is shown in
Fig.~\ref{channel1}(h), and this process will lower E(1B$_u$), which will have
a larger bandwidth upon including all MOs. This last result was also pointed
out in references \onlinecite{Shukla,Ghosh} from transition
dipole moment calculations.

The most important conclusion that one draws from Fig. \ref{channel1} is that
all interactions between minimal basis diagrams and the molecular diagrams
are strictly local. The very nature of the CI process is such that the
configurations that occur in the molecular 2A$_g$ have matrix elements only
with minimal basis TT diagrams in which the two triplet excitations occur on
nearest neighbor units.

\subsection{Two units: Coupling to intermolecular 2e--2h states}

\begin{figure}
\begin{center}
\epsfysize=4cm\epsfbox{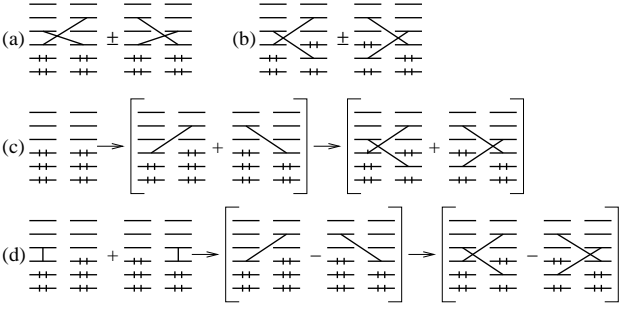}
\caption{(a) and (b): Superpositions of asymmetric
2e--2h exciton basis VB diagrams
with crossed bonds, for the two-unit case. There are many such diagrams and
their superpositions occur in both A$_g$ and B$_u$ subspaces, as is shown
for one of the superpositions in (c) and (d) (see text).}
\label{diagm1}
\end{center}
\end{figure}

Two-unit VB diagrams with crossed bonds can be
classified as symmetric or asymmetric with respect to the mirror plane
between the units.
Both even and odd superpositions of the asymmetric VB diagrams are
then possible, and these VB diagrams therefore occur in both A$_g$ and
B$_u$ subspaces. Examples of such asymmetric VB diagrams involving
MOs beyond the HOMO and LUMO are shown in Fig.~\ref{diagm1}(a) and (b).
The simplest
way to confirm that the superpositions shown do occur in both
subspaces is to apply $H_{1e}^{inter}$ twice on either the Simpson ground
state (for occurrence in the A$_g$ subspace) or the Frenkel exciton state
(for occurrence in the B$_u$ subspace). The application of $H_{1e}^{inter}$
in this multi-MO case is achieved through Eqs. ~\ref{inter_mo} and
~\ref{operators}, 
and the signs originate from the different signs
for each $\psi_{1,k,\sigma}^{\dagger}\psi_{2,k',\sigma}$ or
$\psi_{2,k,\sigma}^{\dagger}\psi_{1,k',\sigma}$.
This is done for the superposition
of Fig.~\ref{diagm1}(a) in Fig.~\ref{diagm1}(c) and (d). The occurrence of
the asymmetric VB diagrams in both A$_g$ and B$_u$ subspaces indicates that
they play no significant role in the 2A$_g$ -- 1B$_u$ crossover, and including
them in our numerical calculations would not have altered the value of $U_c$
in N = 4.

\begin{figure}
\begin{center}
\epsfysize=5cm\epsfbox{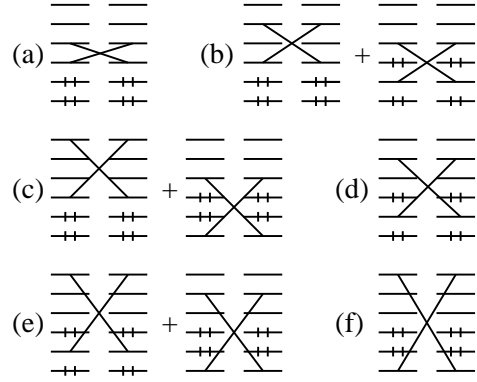}
\caption{Symmetric 2e--2h VB diagrams with crossed bonds
in the two-unit oligomer of the
substituted polyene. (a) is the minimal basis diagram. (b) -- (f): high energy
diagrams ignored in the minimal basis calculation.}
\label{diagm2}
\end{center}
\end{figure}

More important in the present context are therefore the symmetric TT diagrams.
We have shown in Fig.~\ref{diagm2} the corresponding diagrams with
crossed bonds. These symmetric diagrams occur only in the A$_g$ subspace
and can in principle lower the energy of the 2A$_g$.
VB diagram \ref{diagm2}(a) is the lowest energy diagram with crossed bonds,
and we are interested in determining whether this diagram is coupled to any
of the diagrams (b) -- (f) in Fig.~\ref{diagm2}.

\begin{figure}
\begin{center}
\epsfysize=3cm\epsfbox{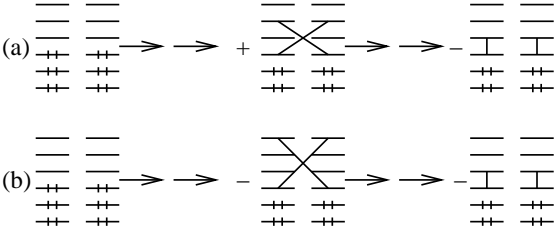}
\caption{CI processes involving two of the four symmetric VB diagrams of
Fig.~\ref{diagm2} that are coupled to the minimal basis symmetric 2e--2h
diagram of Fig.~\ref{diagm2}(a) (see text). The double arrows denote two
consecutive applications of $H_{1e}^{inter}$, promoting interunit CT in
opposite directions. Plus and minus signs are consequences of applying
Eq.~\ref{operators}. Similar CI processes involving the other two symmetric
VB diagrams of Fig.~\ref{diagm2} also give the same final diagram with
minus sign.}
\label{channel}
\end{center}
\end{figure}

Three out of the five sets shown in Fig.~\ref{diagm2},
sets (d), (e) and (f), {\it cannot} have matrix elements of $H_{1e}^{inter}$
with
VB diagram (a) in second order, from orbital occupancies alone.
This is because two applications of $H_{1e}^{inter}$ can change
the occupancies of only two MOs, while the VB diagrams in (d), (e)
and (f) differ from diagram (a) by occupancies of
four orbitals. This
leaves only the diagrams in (b) and (c), which {\it are} coupled to the
diagram in (a) once nonorthogonality of the VB basis functions is taken into
consideration (see Fig.~\ref{basis}(e)), as is shown in the two CI channels in
Fig.~\ref{channel}(a) and (b). The final VB diagram in Fig.~\ref{channel}
has nonzero overlap with diagram (a) in Fig.~\ref{diagm2}, and
hence it might be thought that the CI processes shown can reduce E(2A$_g$).

This is where the signs of the final diagrams in all CI channels become
important. The minus sign in front of the final VB diagram in
Fig.~\ref{channel}(a) and (b), taken together with the overlap of --1/2
between this diagram with noncrossing bonds and diagram \ref{diagm2}(a),
indicates that the diagram \ref{diagm2}(a) is generated with a
{\it plus}
sign from these CI processes. This is in contrast to the CI process in
Fig.~\ref{basis}(a) where diagram \ref{diagm2}(a) is generated with a
minus sign. There is thus destructive interference between the CI channels
and the overall absolute ``yield'' of the low energy diagram with
crossed bonds,
starting
from the Simpson ground state, {\it decreases} when CI channels involving
the two higher energy symmetric diagrams (c) and (e) in Fig.~\ref{diagm2}
are included. Conversely, the relative weight of the Simpson ground state
{\it decreases} in the 2A$_g$, and E(2A$_g$) {\it increases}.
This interpretation of the CI process is unconventional but
is perfectly valid (as can be readily ascertained by constructing
and solving a Hamiltonian matrix that includes all the diagrams and CI channels
shown in Figs~\ref{basis} and \ref{diagm2}, to determine the extent of
mixing between the Simpson ground state and the low energy diagram with
crossed bonds when the interfering channels of Fig.~\ref{channel} are included)
since the Hamiltonian matrix in the exciton VB basis is constructed essentally
from the linear relationships shown in Figs. \ref{basis} and \ref{channel}.

There is an alternate way to arrive at the same conclusion, which uses the 
relationship between VB diagrams in exciton basis and configuration space.
We have already remarked that the singly excited units in the Frenkel exciton
diagrams of Fig.~\ref{basis} are ionic. In the case of polyenes, each 
singly excited unit is the superposition of ionic configurations 20 -- 02,
where the numbers denote atomic site occupancies. Thus two singly excited units
contain (20 -- 02)(20 -- 02), i.e., the ionic configurations 
2020, 2002, 0220 and 0202, and it is only
the superposition in Fig.~\ref{basis}(d) that cancels these ionic 
contributions and gives the covalent VB diagram. Configuration mixing of the
type in Fig.~\ref{channel} therefore involves ionic diagrams at energy
2$U$ (as opposed to covalent diagrams at zero energy), and hence
can only increase E(2A$_g$).

Our overall conclusion then is that already at the level of two units, we
find that interunit high energy diagrams with crossed bonds play a weak
role, and only couplings with intraunit excitations can perhaps lower
E(2A$_g$) [note that such couplings will also lower E(1B$_u$), see
Fig.~\ref{channel1}(h)].
In such a short system, the role of intraunit 2e--2h excitations
can be strong.
If all MOs are included in a FCI calculation, it is possible
that the true 2A$_g$ is below the true 1B$_u$ in N = 4,
although the difference
in their energies should be considerably smaller than in butadiene.
Conversely though, if the molecular unit is such that the molecular 2A$_g$
is {\it higher} in energy than the molecular 1B$_u$,
our discussion indicates that
the 2A$_g$ in the coupled system must necessarily be higher in energy than the
corresponding 1B$_u$. This is in agreement with our previous calculations 
on 
oligomers of PDPA \cite{Shukla,Ghosh} that used a 
multiple reference double-CI approach using a completely delocalized MO basis,
and that found that the oligomer 2A$_g$ is above the oligomer 1B$_u$. 
The unit in PDPA is trans-stilbene, in which the molecular 2A$_g$ is higher than
the molecular 1B$_u$.

\subsection{Longer chains: Coupling to interunit 2e--2h states}

For longer chains
we focus on interunit 2e--2h VB diagrams only, since all interactions
involving molecular 2e--2h states are restricted to minimal basis diagrams with
nearest neighbor TT bonds.
As discussed in the above,
CI processes of the type shown in Fig. \ref{CI}(b) are particularly important,
because of the degeneracy between the short and long-bonded crossed diagrams.
The nature of the exact 2A$_g$ here and in reference \onlinecite{Chandross}
both indicate this.
One can then visualize such CI processes involving all pairs of 
MOs $k, k'$, which generate diagrams with long crossed bonds in longer chains,
with the relative weights of the diagrams with short crossed bonds spanning
two units gradually decreasing with chain length. 
Whether or not these diagrams with long crossed
bonds are coupled to the minimal basis TT diagrams then decide how strongly
the calculated minimal basis $U_c$ is affected.

\begin{figure}
\begin{center}
\epsfysize=4cm\epsfbox{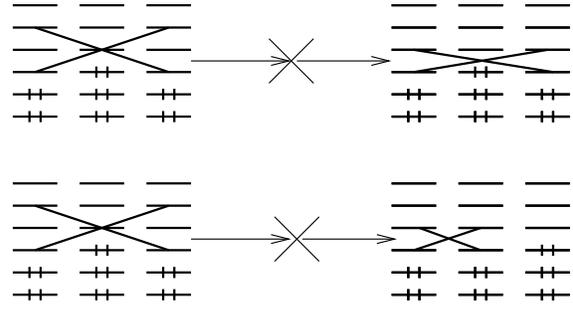}
\caption{The absence of coupling in second order between high energy
long bonded 2e--2h
VB diagram with crossed bonds and minimal basis VB diagrams with crossed bonds.
Two applications of $H_{1e}^{inter}$ on the diagram on the left does not 
generate either of the two diagrams on the right.}
\label{longbond}
\end{center}
\end{figure}

The classification
of intermolecular 2e--2h VB diagrams as mirror plane symmetric or asymmetric
persists in longer chains. Consider first the class of symmetric VB diagrams
with long crossed bonds between distant units. Between any pair of molecular
units there are as many of these as in Fig.~\ref{diagm2}.
Since, however, $H_{1e}^{inter}$ consists
of nearest neighbor CT only, there are {\it no} symmetric diagrams with long
bonds that are coupled to the lowest energy diagram with crossed bonds in
second order now.
This is shown in Fig.~\ref{longbond}, where we have taken
a three-unit version of one of the diagrams in Fig.~\ref{diagm2} that did
have coupling with the lowest energy diagram with crossed bonds. {\it Thus the
minimal requirement of having the same occupancy in two out of four MOs
is no longer sufficient to give coupling to the low energy diagram.}

\begin{figure}
\begin{center}
\epsfysize=2.8cm\epsfbox{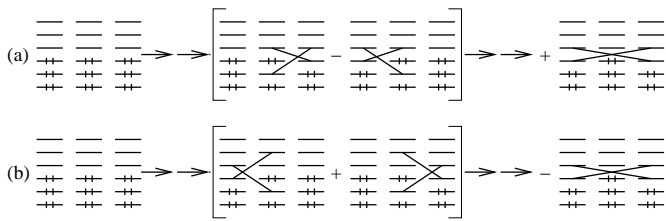}
\caption{Examples of CI processes that can generate the long bonded minimal
basis VB diagram with crossed bonds, starting from the Simpson ground state,
with high energy asymmetric 2e--2h diagrams with crossed bonds as the
intermediate states.
The double arrows again denote two consecutive applications of 
$H_{1e}^{inter}$. Note the opposite signs for the final product in the
two channels, which is a consequence of charge-conjugation symmetry.}
\label{antisymm}
\end{center}
\end{figure}

New CI channels become important now with the asymmetric diagrams with
crossed bonds.
A subset of the asymmetric diagrams
can now give long-bonded minimal basis TT diagrams, as
shown in Figs ~\ref{antisymm}(a) and (b), where we have given two such
examples. {\it It should be clear that the reverse of what is shown in this
Figure, viz., generation of short-bonded minimal basis
TT diagrams from long-bonded asymmetric diagrams is equally probable.}
In general high energy asymmetric diagrams with bonds spanning L units
can give low energy symmetric diagrams with crossed bonds spanning L -- 1
or L + 1 units.
We discuss only the former, since the essential physics of all these processes
are identical, and originate from charge-conjugation symmetry alone. Since
we are interested in two consecutive CT processes in opposite directions only,
the particular subset of asymmetric diagrams with short crossed bonds that
can give the long bonded minimal basis TT diagram must
satisfy a very strict condition, viz.,
{\it both the HOMO and the LUMO of one of the molecules in the initial
state must be singly occupied.}
For a reason that will be clear below, we have given
in Fig. ~\ref{antisymm}
the complete CT steps,
starting from the Simpson ground state, that show how the asymmetric VB
diagrams are generated in second order, and then another second order process
gives the minimal basis VB diagram with long crossed bonds. 
The CI between the high energy asymmetric 2e-2h diagrams and the minimal
basis diagram with crossed bonds clearly seems to suggest that the former are
relevant, according to the two criteria of relevance we have proposed,
and can lower E(2A$_g$).

There are two reasons why CI processes such as those in Figs.~\ref{antisymm}
(and all similar processes in which asymmetric diagrams with crossed bonds
of length L generate symmetric diagrams with minimal basis crossed bonds with
lengths L+1 or L--1)
contribute very weakly to the exact 2A$_g$. First,
the intermediate VB diagram with short crossed bonds
in Fig. ~\ref{CI}(b) is degenerate with the final TT state (as already
pointed out in the above, this is what makes the contribution of long bonded
TT diagrams in the 2A$_g$ of polyenes large), while both the
intermediate 2e--2h states in Fig. ~\ref{antisymm} have diagonal energies that
are different from that of the final TT state. Even more importantly, we note
that while the ``yield'' of the long bonded crossed diagram in
Fig.~\ref{antisymm}(a) is positive, that in Fig.~\ref{antisymm}(b) is
negative.
{\it The opposite signs result from straightforward applications of
Eq.~(\ref{operators}) and are a consequence entirely of 
charge-conjugation symmetry, which determines the signs in these equations.}
This has a very important implication for the extent of configuration
mixing between the Simpson ground state and the long bonded TT diagram,
viz., the configuration mixing induced by any one channel is reduced
by the other. The
simplest way to see this is to imagine that the two superpositions of
asymmetric diagrams in Fig.~\ref{antisymm}(a) and (b) 
have the same diagonal matrix elements
of $H_{ee} + H_{1e}^{intra}$. In that case, by operating with $H_{1e}^{inter}$
we would have simply obtained
the superposition of all four diagrams (the relative signs between which are
determined entirely by Eq.~(\ref{operators})), and now a second application of
$H_{1e}^{inter}$ would simply give zero.

\begin{figure}
\begin{center}
\epsfysize=3.5cm\epsfbox{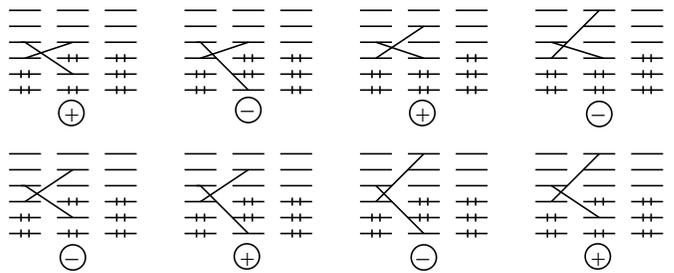}
\caption{Complete list of all asymmetric 2e--2h high energy VB diagrams
with crossed bonds that are coupled to the Simpson ground state as well
as the long bonded minimal basis VB diagram of Fig.~\ref{antisymm}, through
CI processes of the type shown there. The plus or minus sign associated
with each diagram corresponds to the sign on the minimal basis diagram,
when the CI channels as in Fig.~\ref{antisymm} are constructed, starting
from the Simpson ground state in each case.}
\label{asymm2}
\end{center}
\end{figure}

In Fig.~\ref{asymm2} we have given all the asymmetric diagrams with
crossed bonds in the
three-unit chain that can be obtained by applying $H_{1e}^{inter}$ twice
to the Simpson ground state, and that in turn give the long-bonded 
minimal basis diagram with crossed bonds
when further operated on by $H_{1e}^{inter}$ (in each case we have given one
diagram only, but superpositions of the type in Fig.~\ref{antisymm} are
being implied). The positive and negative sign
associated with each diagram is the sign on the long bonded final diagram when
CI channels as in Fig.~\ref{antisymm} are constructed with these diagrams
as intermediate states. Not surprisingly, the total number of such diagrams
in Fig.~\ref{asymm2} is even. Taken together with the fact that the
configuration mixing between the Simpson ground state and the long-bonded
TT diagram due to any one of these channels is small to begin with
(due to the nondegeneracy between
the intermediate asymmetric state and the final state), the even number
of channels, with half of them giving positive matrix elements and the other
half giving negative implies a nearly total cancellation effect. Since the
above result is a consequence of charge-conjugation symmetry alone, it remains
valid for all sets of asymmetric diagrams with crossed bonds, independent
of the bond lengths. We therefore
conclude that {\it exactly as the symmetric diagrams with long crossed
bonds, the
asymmetric diagrams with crossed bonds have very weak
CI with the minimal
basis TT diagrams in long chains.}

\subsection{Summary}

Our basic conclusion is that high energy intraunit 2e--2h
excitations that have been ignored in our minimal basis calculations can
affect the final results. However, couplings between these and minimal basis
TT diagrams are strictly local, and
moreover, due to the nondegeneracy between the molecular 2e--2h states
and the minimal basis 2e--2h states the effect of this coupling on long chains
should be weak. 
Recall, for example, that the relative weights of minimal basis TT
diagrams with nonnearest neighbor bonds can continue to be relatively 
large even as the lengths of the interunit bonds keep increasing
(see Fig.~\ref{2Ag} and also reference \onlinecite{Chandross}),
but the couplings in Fig.~\ref{channel1} are only to TT diagrams with nearest
neighbor bonds.
In contrast to the intraunit 2e--2h states, the coupling of the minimal
basis TT diagrams to the interunit high energy
2e--2h VB diagrams is extremely weak. Indeed this latter coupling can even
raise E(2A$_g$) slightly in the case of two units. 
The weak coupling with high energy interunit states
is a symmetry effect that will persist in any system which has spatial
and charge-conjugation symmetries.

\section{Conclusion}
\label{conclusion}
In conclusion, we have analyzed the excited state ordering in finite chain
analogs of the simplest model conjugated polymer with transverse
$\pi$-conjugation within the dimerized Hubbard Hamiltonian. The system
considered consists of a backbone polyene chain with ethylenic
sidegroups. We developed a
multilevel exciton basis description for the system, and within a minimal
basis, which consists of the HOMO and the LUMO of each unit, we showed that
the effective electron correlations in this substituted polyene is smaller
than that in the unsubstituted polyene. The reason for this is that each
carbon atom of the polyene is now replaced with multiple atoms over which
a double occupancy is delocalized, thereby giving it a transverse bandwidth
that lowers its energy \cite{Shukla}. Smaller effective correlations, in turn,
suggests that the 2A$_g$ is above the 1B$_u$ in this class of materials, and
that this provides an interesting route to obtain organic polymers that emit
in the IR.

The chief advantage of the exciton
basis approach is that the role of the high energy configurations
that are ignored
within the minimal basis calculations
can be determined precisely, because of the local character of the basis
functions, as well as the local nature of the configuration interaction within
the Hubbard Hamiltonian. Our very detailed analysis here indicates that few
of the high energy TT configurations
that we ignore have CI in second order with the
minimal basis TT configurations. 
In particular, the CI between interunit high and
low energy TT diagrams is very weak. The only relevant CI between high 
and low energy states that can at all lower the energy of the polymer 2A$_g$  
involves the intraunit configurations that describe the molecular
2A$_g$ state. Note that this already ensures the following result, viz., if the
molecular unit is such that the molecular 2A$_g$ is higher than the
molecular 1B$_u$, then since within the minimal basis the 2A$_g$ is also
higher than the 1B$_u$, the overall E(2A$_g$) is also higher than the overall
1B$_u$. This in turn predicts that that E(2A$_g$) must necessarily be 
larger than E(1B$_u$) in PDPA, since the molecular 2A$_g$ is higher than the
molecular 1B$_u$ in the fundamental unit trans-stilbene.  
Our conclusion here is in agreement with the
previous conclusion about PDPA, which
was arrived from computations only \cite{Shukla,Ghosh}. 

The molecular 2A$_g$ configurations 
interact only with the minimal basis
TT diagrams in which the two triplets occupy nearest neighbor units. In a long
chain system, the relative weight of such TT diagrams with nearest neighbor
bonds is small, since there exist in long chains TT diagrams with all
possible bond lengths, and their contributions are comparable.
At the same time,
in the isolated real 
hexatriene molecule, which forms the unit in our model system,
the lowest one- and two-photon states occur at a very high energy of
$\sim$ 5 eV, whereas the
polymeric 1B$_u$ and 2A$_g$ occur below 1.6 eV, which is 
E(1B$_u$) for t-PA. Thus any interaction between molecular and polymer states
decreases progressively with increasing chain length, and we expect
that even for our model system  
the minimal basis results are qualitatively correct. 
Importantly, of course, our model calculations are for the illustration of
the principle only, and real systems in which the molecular 2A$_g$ is higher 
than the molecular 1B$_u$ can always be developed with suitable
modification of the molecular unit. The real experimental challenge here
is to develop systems in which steric interactions are small enough that 
oligomers with long conjugation lengths and small optical gaps can be
synthesized. Finally, we mention that we have not discussed here known 
$\pi$-conjugated polymers with low band gaps, such as
poly(isothianaphthene) \cite{Kobayashi}
and poly(isonaphthothiophene) \cite{Ikenoue}. It is of interest that
in these systems too there occur transverse conjugation over a few bonds in
addition to the conjugation along the polymer axis, although the nature of
the transverse conjugation is slightly different from that in our model system
consisting of a substituted polyene. The electronic structures and the small
optical gaps in these systems have been discussed within one-electron theory
\cite{Kertesz} and future theoretical work within many-electron approaches 
is clearly of interest.

\section{Acknowledgments}

Work in Arizona was supported by NSF DMR-0101659, NSF ECS-0108696 and by the 
ONR.
Sandia is a multiprogram laboratory operated by Sandia Corporation, a
Lockheed Martin Company, for the United States Department of Energy under
Contract DE-AC04-94AL85000.

\bibliography{co}
\end{document}